# Preparation of isotopically enriched $^{112,116,120,124}$Sn targets at VECC

*Ratnesh Pandey[1*], S.Kundu[1,2], K.Banerjee[1,2], C. Bhattacharya[1,2], T.K.Rana[1,2], G.Mukherjee[1,2], S.Manna[1,2], J.K.Sahoo[1], H.Pai[3],T.K.Ghosh[1,2], Pratap Roy[1,2], A.Sen[1,2], R.S.M. Saha[1], J.K.Meena[1], A.Saha[1],D.Pandit[1,2], A.Datta[1,2]*

[1]*Variable Energy Cyclotron Centre,1/AF, Bidhan nagar, Kolkata, 700064*
[2]*Homi Bhabha National Institute, Training School Complex, Anushakti Nagar, Mumbai - 400094, India*
[3]*Saha Institute of Nuclear Physics, 1/AF bidhan Nagar, Kolkata, 700064*

**ABSTRACT:**

Resistive heating and mechanical rolling methods have been employed to prepare isotopically enriched thin target foils of $^{116}$Sn (~380μg/cm$^2$), $^{124}$Sn(~400μg/cm$^2$) and thicker foils of $^{112}$Sn (1.7 mg/cm$^2$),$^{120}$Sn (1.6 mg/cm$^2$),respectively. Preparation of enriched targets with small amount of material, selection of releasing agent for thin targets and separation of deposited material insolvent were among the several challenges while fabrication of the thin targets. Uniformity of the targets has been measured using $^{241}$Am α-source. NaCl has been used as releasing agent in preparation of the thin targets. These targets have been successfully used in nuclear physics experiments at VECC.

## 1. Introduction:

Target foils play an important role in the low-energy, heavy-ion induced nuclear reaction experiments. Extensive studies have been performed by us to understand the fragment emission mechanisms for low-energy nucleus-nucleus collisions. These studies revealed that, for low energy (10 MeV/u), light heavy-ion ($A_{proj}$ + $A_{target}$<60) collisions, fusion followed by asymmetric fission (FF) [1] is the dominant mechanism which contributes to the observed fully energy damped yields of the fragments. However, the situation becomes complex for the reactions involving α-cluster nuclei, where nuclear structure is also known to play an important role in the equilibrium emission of complex fragments. In addition to the standard fusion-fission route of fragment emission, the projectile and the target have a finite probability to form a long-lived dinuclear composite [2,3,4], which directly undergoes scission (without the formation of the fully equilibrated compound nucleus) to emit complex fragments. It has been observed that deep inelastic orbiting mechanism plays a significant role in fragment emission from the reactions involving α-cluster nuclei (e.g., $^{20}$Ne + $^{12}$C [2], $^{24}$Mg + $^{12}$C [5], $^{28}$Si + $^{12}$C [6], etc.). As a matter of fact, target foils play a vital role for such type of studies as the energy resolution is a major concern for such type of experiments.

In nuclear reaction and structure studies target materials are bombarded with high energy particles. This requires an isotopically and chemically pure target material prepared as a self supporting thin film, or as alternative, prepared on a backing materials. The target materials are in metallic or gaseous form. The deposited film thickness must be adjusted to the respective experiment under study while observing the optimum thickness. The targets for nuclear physics experiment can be prepared by different methods such as physically from vacuum vapor deposition [7-20], chemically by electrodeposition [21] from aqueous media and molecular plating from organic solvent or centrifugal method [22], mechanically by rolling [23-24] of sheets, etc. A thin target offers different advantages by reducing the multiple scattering, spread due to straggling within the target and enhancing the probability of the particles getting out of the target. On the other hand thicker targets are required for significant reaction cross section. In this paper the preparation of self supporting $^{112,116,120,124}$Sn (thin $^{116,124}$Sn and thick $^{112,120}$Sn) targets by resistive heating and mechanical rolling method has been discussed in details. Isotopic Sn targets have been prepared with the motivation to study the fragment emission mechanism and effect of iso-scaling in $^{16}$O+$^{112,124}$Sn and $^{20}$Ne+$^{112,116,124}$Sn reactions. In beam experiments have been performed at VECC with $^{16}$O and $^{20}$Ne beams. The paper has been arranged in the following manner: fabrication of thin isotopic Sn targets are discussed in section2, preparation of thicker isotopically enriched Sn targets are discussed in section 3, uniformity and purity measurement of the targets are presented in section 4, experiments using isotopic Sn targets are discussed in section 5 and finally conclusion is given in section 6.

## 2. Preparation of thin targets:

Highly uniform, thin Sn targets cannot be prepared by rolling. Since the young modulus of Sn is ~50Gpa, so it is a soft material in nature and it is difficult to prepare its thin target by rolling. Thin targets of isotopic Sn material have been prepared by high vacuum evaporation [10-12] method. Due to difficulties in fabricating such targets many groups have preferred deposition of Sn on backing. Sn targets on carbon [10] and Aluminum[11 ] backings have been reported in literature. Recently, Arshiya et.al [12] has prepared thin[116,118]Sn targets by vacuum evaporation technique. Isotopically enriched materials are very expensive because of their poor abundance in nature. Therefore, care must be taken for the optimum use of material for the preparation of target and accordingly the methods are also chosen in which the loss of material is minimum. Natural abundance of Sn isotopes is given inTable.1. Comparison of present work with others reported work is shown in Table.2. It is clear from the table that self supporting $^{120}$Sn and $^{124}$Sn targets are not yet reported in literature.

Table.1. Abundance of Sn isotopes.

| Isotopes | $^{112}$Sn | $^{114}$Sn | $^{115}$Sn | $^{116}$Sn | $^{117}$Sn | $^{118}$Sn | $^{119}$Sn | $^{120}$Sn | $^{122}$Sn | $^{124}$Sn |
|---|---|---|---|---|---|---|---|---|---|---|
| Abundance (%) | 0.97 | 0.66 | 0.34 | 14.54 | 7.68 | 24.22 | 8.59 | 32.58 | 4.63 | 5.79 |

Table.2. Comparison of present work with other reported works.

| Isotopic targets | Thickness | Preparation method |
|---|---|---|
| $^{112}$Sn | 1.7 mg/cm$^2$ self supporting (**Present work**)<br><br>2.5 mg/cm$^2$ on Pb backing Ref.[23]<br><br>1 mg/cm$^2$ self supporting Ref.[24] | Rolling of ingot |
| $^{116}$Sn | ~380 μg/cm$^2$(**Present work**)<br><br>~250-600 μg/cm$^2$self supporting Ref.[12]<br><br>~150μg/cm$^2$ on carbon backing of ~25μg/cm$^2$. Ref.[10] | Resistive heating |
| $^{118}$Sn | ~250-600 μg/cm$^2$ self supporting Ref. 12] | Resistive heating |
| $^{120}$Sn | 1.6mg/cm$^2$ self supporting (**Present work**) | Rolling of ingot |
| $^{122}$Sn | (8.4mg/cm$^2$) self supporting Ref.[25] | Rolling of ingot |
| $^{124}$Sn | ~400μg/cm$^2$ self supporting(**Present work**)<br><br>~200-350μg/cm$^2$on Al backing of (~1.2-1.7 mg/cm$^2$). Ref. [11] | Resistive heating |

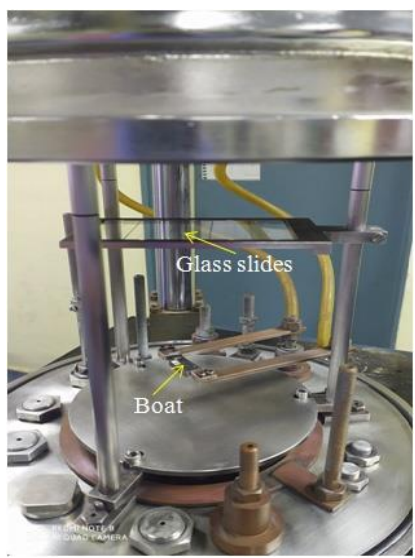

Figure 1. Target preparation setup.

**2.1 Evaporation of chemical salt:**

Several test runs have been made with natural Sn material before the final deposition with enriched material to optimize the process with least material. After optimization of fabrication method with natural material, isotopically enriched $^{116,124}$Sn have been deposited by thermal heating. Thin isotopic targets were prepared by resistive heating technique using dome shaped coating unit shown in Figure 1. Chamber was cleaned gently with scrubber and wiped with alcohol dipped tissue paper. The important thing to be noted in target preparation is the choice of boat which has been used for making of the targets. Smaller the size of the boat, bigger the solid angle coverage for evaporation. Therefore, a small square shape Tantalum (Ta) boat was used for target preparation after several successful attempts with natural material. Tantalum boat was fixed in between copper plates as shown in Figure 1 and glass slides were kept at ~10cm above the boat. After fixing the boat and slides inside the chamber, the dome of the chamber was kept closed for vacuum. Rotary pump and diffusion pump were used for achieving high vacuum (~$10^{-6}$ mbar) inside the chamber of the coating unit. Several attempts have been made to choose suitable releasing agent for Sn target. Teepol solution and different chemical salts were tested as releasing agent. Arshiya et.al [12] has reported KCl as releasing agent for preparation of isotopic tin targets. Since NaCl is more soluble then KCl in water, so material deposited on sodium chloride will remove quickly than it deposited on potassium chloride. It was found that NaCl was best choice of releasing agent for self supporting Sn targets. 13 mm diameter pellet of NaCl salt were prepared by Hydraulic press method using pellet maker. The prepared pellet was kept in Ta boat of size (27x12) mm as shown in Figure 2(a) for evaporation in close chamber. Rotary and diffusion pumps were used for achieving high vacuum. After achieving high vacuum of the order of ~$8x10^{-6}$ mbar, heating process was started. Current of coils was slowly increased for homogeneous coating of salt on glass slide. Initially it was kept ~30A for 5 minutes, to reduce the moister inside the chamber. Coil current was slowly increased up to ~100A for evaporation of NaCl pellet. Change of state of salt from solid to liquid and liquid to vapor were started and vapors of salt was deposited on glass slides. After the deposition of salt on slides, the coil current was slowly reduced to zero and chamber was kept closed for few hours for cooling.

**2.2 Thermal evaporation of enriched materials:**

Since Sn has the low melting point ~232°C, therefore resistive heating technique was used for preparation of thin $^{116,124}$Sn targets. The weight of enriched material was taken before the deposition. The chamber was opened for few minutes to load the material inside small square shaped boat of the size (7x7) mm as shown in Figure 2(b). Precautions were taken to keep the material inside the boat. This was done very rapidly to avoid the moisture entering inside the chamber which would affect on the releasing agent deposited on the substrate. Isotopic material of $^{116}$Sn (70 mg) of

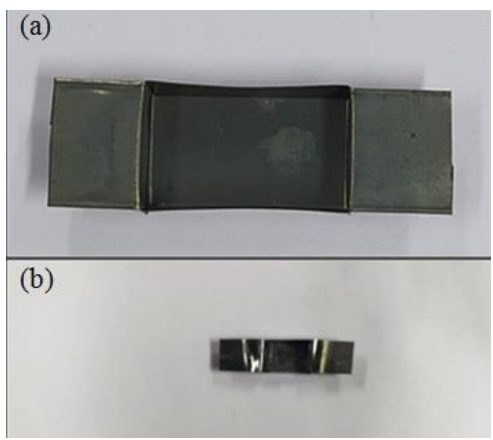

Figure 2. Tantalum boats used in thermal heating of Sn targets (a) Boat for salt. (b) Boat for isotopic material.

~98% purity and $^{124}$Sn (77mg) of ~99.9% purity has been used for evaporation. Within few minutes the material was loaded on Ta boat and chamber was closed for vacuum. The typical distance between boat and substrate was optimized after several runs and it was kept around ~8cm. Photograph of setup for preparation of thin targets is shown in Figure 1. When vacuum of ~$10^{-6}$ mbar was achieved, evaporation coil current was slowly increased and at I~120mA, the powered material was converted in form of bubble and thereafter its state changed from liquid to vapor. As the evaporation process started, the vacuum of the chamber started decreasing due to movement of molecules inside the chamber. The stopwatch was kept on during deposition, in order to note down deposition time. The typical deposition time for $^{116}$Sn target was 11.26 minutes whereas for $^{124}$Sn it was 12.32 minutes. After whole material was evaporated, the current of coil was slowly moved to zero.

## 2.3 Mounting of the isotopic target materials:

The glass slide coated with substrate and material was inserted into a vessel with a solvent appropriate to the releasing agent deposited on it. De-ionized water was taken as a solvent for enriched Sn targets. Before releasing the deposited material in solvent, a square cut mark bigger than the size of target holder was made on substrate in order to separate the individual targets from the deposited layer. The slide was slowly inserted inside glass container with an angle ~$30^0$-$45^0$. Care has been taken while releasing the thin film from substrate to solvent. The release of the target material from substrate to solvent was preceded by raising the solvent level at proper speed. The solvent level can be raised by slow immersing of substrate into the vessel. When a thin film is released from the substrate to solvent it can be mounted on target frame by fishing it from the solvent surface. The slides were immersed perpendicularly to the foil and the solvent surface and kept in this position when lifting up the foil to minimize the impact of the solvent surface tension. Photograph of enriched targets floating in glass vessel is shown in Figure 3. After the target dried, the self-adhesive forces are sufficient to hold the foil in place in case of thin foils, but for thicker ones the application of appropriate glue or clamping the foil between two frames may be needed. Figure 4 shows isotopically enriched Sn thin film mounted on 13mm diameter target holder. The prepared targets have been successfully used in online experiments at VECC. A dark spot of beam is clearly visible on targets used in experiments, Figure 5 (a) and (b). Preparation of $^{124}$Sn target is not yet reported so for.

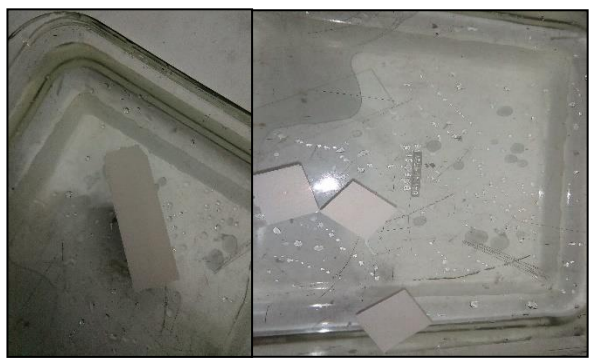
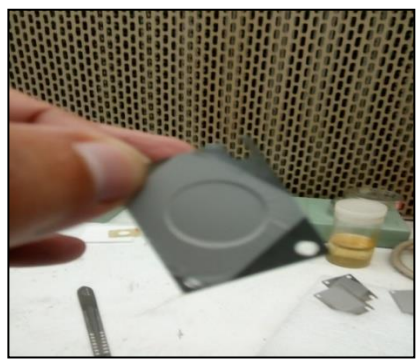

Figure 3. Floated thin targets in glass vessel.                    Figure 4. Mounting of target on target holder.

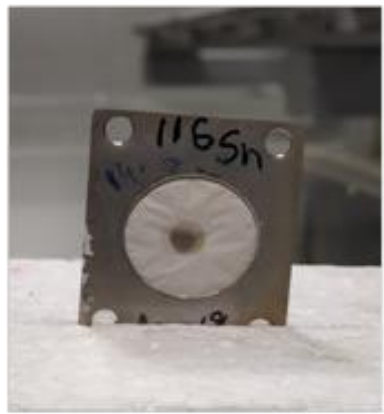
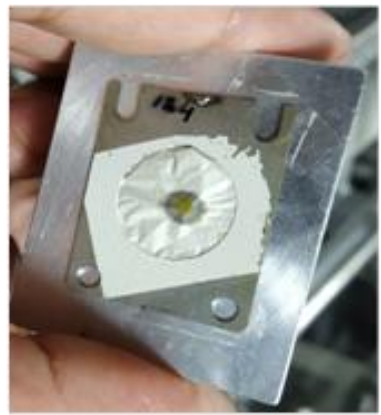

Figure 5.(a) $^{116}$Sn target (b) $^{124}$Sn targets used in online experiments at VECC.

## 3. Preparation of $^{112,120}$Sn targets by rolling method:

This technique should be considered as one of the primary methods of target preparation, particularly for expensive isotopically enriched materials as the waste of material with this technique is negligible and the method is known for preparing a high quality targets. In this method the material is reshaped (transformed into foil) using a rolling machine. Self-supporting isotopic Sn targets of uniform thickness were needed for experiments at VECC. Thicker targets were prepared via rolling technique [23-25], however generally it is assumed that the lower thickness limit for most of the metals are at the level of 1 mg/cm$^2$. The achievable thickness of the final product varies a lot from material to material depending on its malleability. Several attempts have been made for preparation of self-supporting Sn targets using natural powdered tin but it was difficult to prepare self-supporting isotopic Sn target using powered Sn material. Recently thicker Pb backed $^{112}$Sn target has been prepared using rolling of ingot method [23]. We have followed the same process for preparation of powder form of material to solid ingot. Isotopic $^{112,120}$Sn of ~99.6% and 98.8% purity has been used in powered form; weight of the material for preparation of ingot was ~10 mg. The ingot was rolled with rolling machine shown in Figure 6 at VECC. Care was taken while rolling of soft Sn material because after attaining the lower thickness limit, the material sticks on the stainless steel plate. The thickness of prepared $^{112}$Sn & $^{120}$Sn targets are 1.7 mg/cm$^2$ and 1.6 mg/cm$^2$ respectively. Figure 7(a) and (b) shows

[112,120]Sn targets respectively used in VECC experiment. We have achieved better thickness isotopic [112]Sn targets using less amount of material. [120]Sn targets has not yet prepared via rolling of ingot. The thickness of prepared [112,116,120,124]Sn targets have been measured by energy loss method [26] using [229]Th alpha-source in laboratory. The method is based on the energy loss of alpha particle passing through the target material.

## 4. Uniformity and purity measurements:

### 4.1 Uniformity measurement of isotopic targets:

Uniformity of [116]Sn target have been measured using [241]Am (α, pin source). Energy spectrum at five positions, i.e centre, centre right, centre left, center up and centre down have been taken and it was found that energy loss for each positions were same. This confirms that thickness is uniform throughout the surface of target. Figure 8 shows the energy spectrum by α source taken at different positions of target, In section (a) energy spectrum have been taken with source and without target. Similarly Sections (b), (c), (d), (e) and (f) have been taken at centre, center left corner, center right corner, center up and center down positions, respectively. It is clearly visible from energy spectrum that for all positions energy spectrum is same which confirms the similar energy loss at all positions of target. Similarly all targets have uniform thickness through the surface of target.

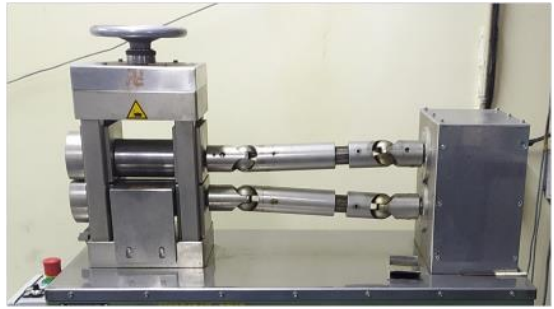

Figure 6: Mechanical rolling machine for target preparation at VECC.

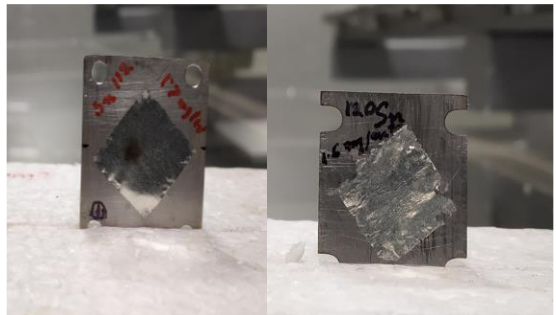

Figure 7: (a) [112]Sn target (b) [120]Sn target prepared by mechanical rolling method.

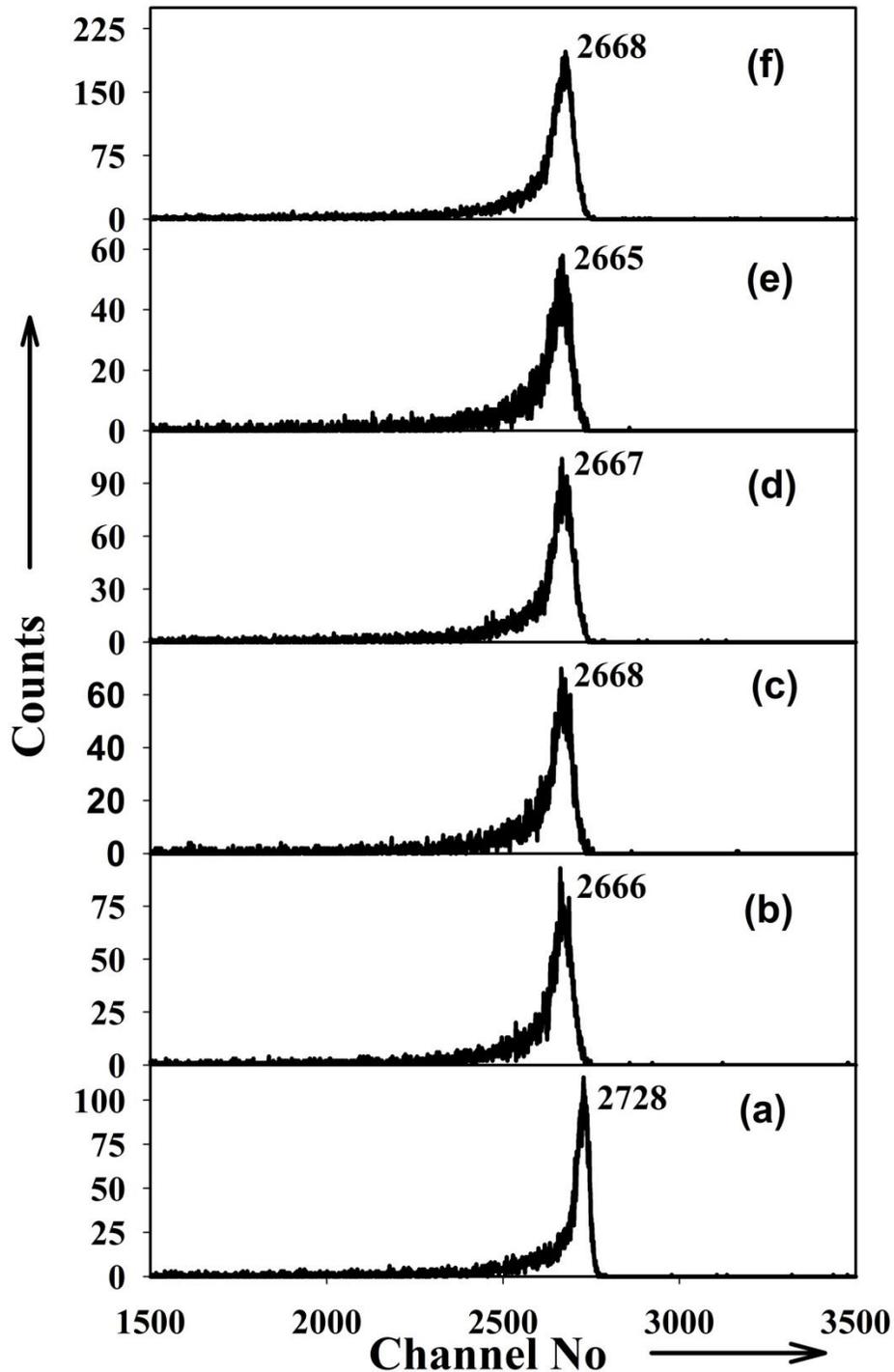

Figure 8: Energy spectrum obtained by Si detector (a) blank, (b)centre position of target, (c) centre left corner, (d) center right corner, (e) center up and (f) center down corner.

## 4.2. Energy dispersive X ray spectroscopy (EDS) measurement

Elemental compositions of isotopic targets were identified by energy dissipative X ray spectroscopy (EDS). The EDS measurements of prepared thin targets have been performed using Carl Zeiss make Supra-55 FESEM with Oxford make EDS detector. Elemental analysis for $^{112,116,120,124}$Sn targets is shown in Figure. 9. It is clearly observed from the figure that weight% of 99.7%, 99.7%,95.5% and 98.7% of Sn is present, However weight% of present impurities are 0.3% Ti,0.3% Ti, 4.5% O and 1.3% Al for $^{112}$Sn, $^{116}$Sn, $^{120}$Sn and $^{124}$Sn targets, respectively. This result shows that other impurities are negligible in all enriched targets.

## 4.3 Surface morphology

Advanced characterization technique such as field emission scanning electron microscopy (FE-SEM) is used for surface morphology of target materials. The FE-SEM images were obtained using Carl Zeiss make Supra-55 FESEM with Oxford make EDS detector at 10KV, AC Voltage. Morphological changes are clearly visible in targets prepared by vacuum evaporation and rolling methods, as shown in Figure 10. It is clear that roughness of the film increases with decrease in thickness. From SEM images we can also predict that resolution is higher in thinner films as compared to thicker targets.

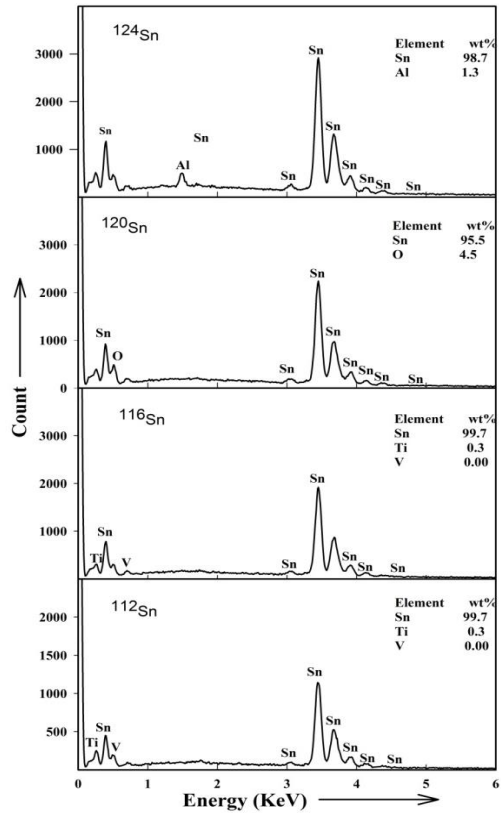

Figure 9: Energy Dispersive X-ray Spectroscopy (EDS) of Isotopically enriched $^{112,116,120,124}$Sn targets.

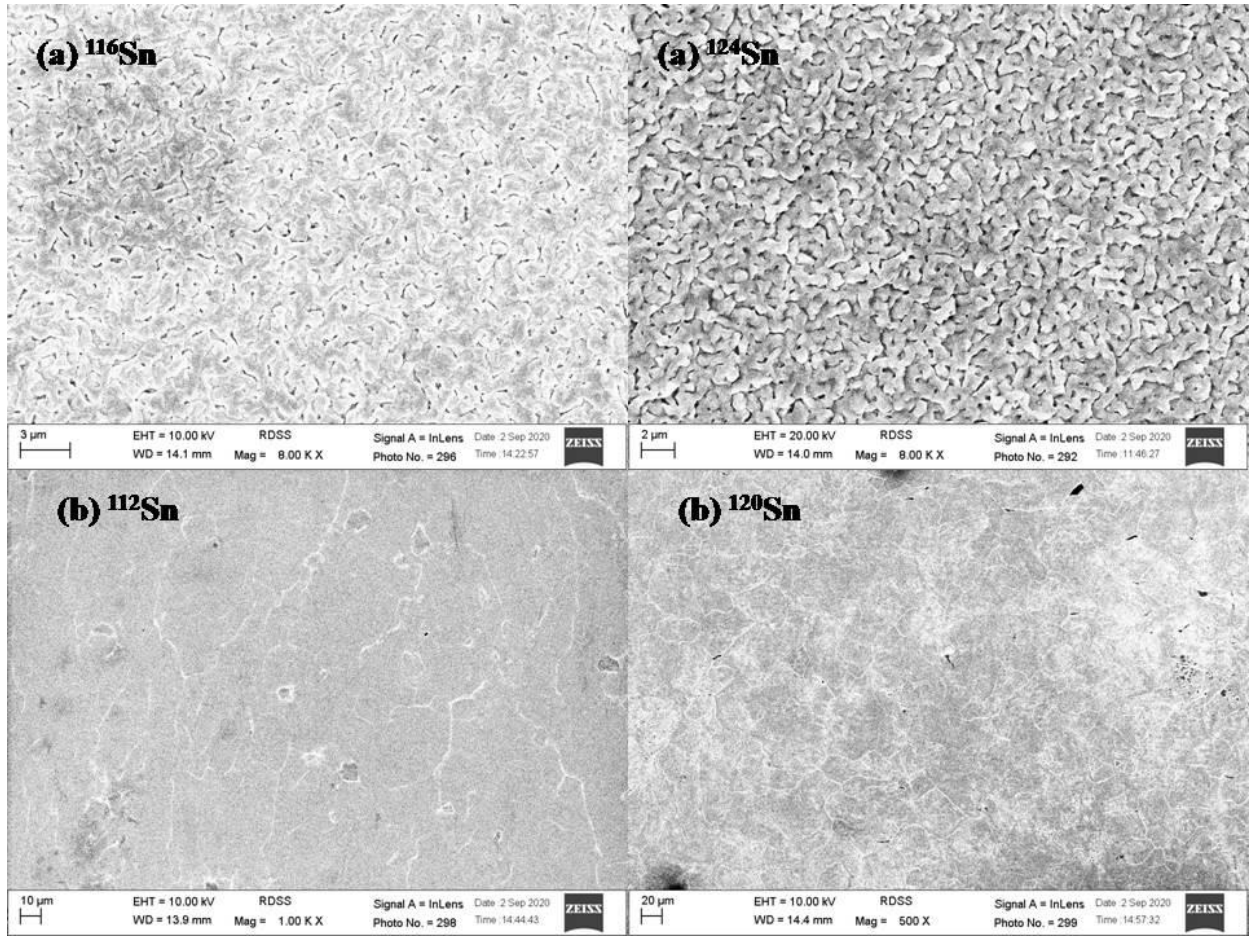

Figure 10: FE-SEM images of (a) $^{116}$Sn and $^{124}$Sn (b) $^{112}$Sn and $^{120}$Sn targets.

## 5. Experiment using isotopic Sn targets:

Two experiments were performed using $^{112,116,120,124}$Sn targets with $^{16}$O and $^{20}$Ne beams from K130 Cyclotron at VECC. The experiments have been carried out using $^{16}$O (120, 156 MeV, 126.4, 159.7 MeV) and $^{20}$Ne (160MeV and 157.4MeV) beams on $^{112,116,120,124}$Sn targets respectively. The motivation of experiments was to study fragment emission mechanism and isoscaling effect. Three telescopes were used for detection of fragments (2<Z<7). Telescope one was single sided ~50-μm-thick Si (ΔE) strip detector followed by a double sided ~525-μm-thick Si(E) strip detector. Telescope two was~50 μm-thick Si (ΔE) detector, followed by CsI(Tl) (E) detector and telescope three was ~100 μm-thick Si (ΔE) detector, followed by CsI(Tl)(E) detector. The isotopic separations obtained in these experiments are illustrated by the *ΔE* vs *E* plot displayed in Figures 11 and12 for the reactions$^{16}$O+$^{124}$Sn and $^{20}$Ne+$^{112}$Sn respectively. Well separated ridges are clearly seen corresponding to elements having atomic numbers up to *Z* = 8 and clear isotopic separation have been obtained for the fragments up to *Z* = 6.

## Conclusion

Isotopic metal films of Sn from lowest to highest isotopes having different thicknesses have been fabricated using vacuum evaporation and rolling techniques. Uniformity of the targets has been measured using α source at different positions of targets. Purity and surface study of these targets have been characterized by EDS and FESEM techniques. It was found that there were negligible impurities in enriched targets. The enriched Sn targets were used in online experiments using $^{16}$O and $^{20}$Ne ion beams at VECC.

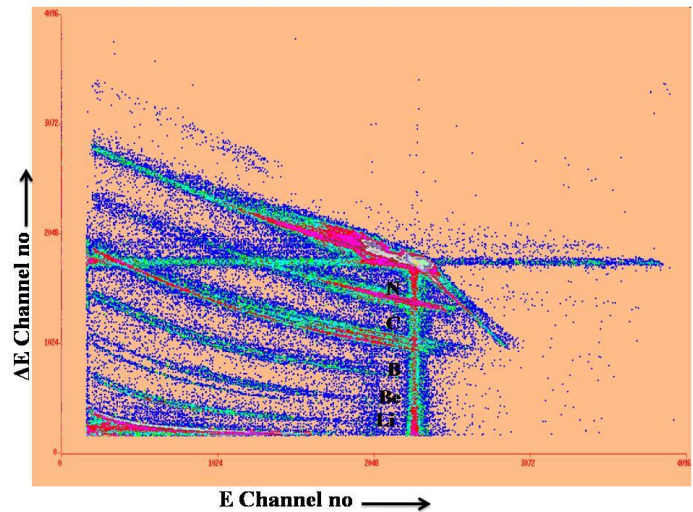

Figure 11: Typical two dimensional spectrum of fragments obtained in the reaction $^{16}$O+$^{124}$Sn.

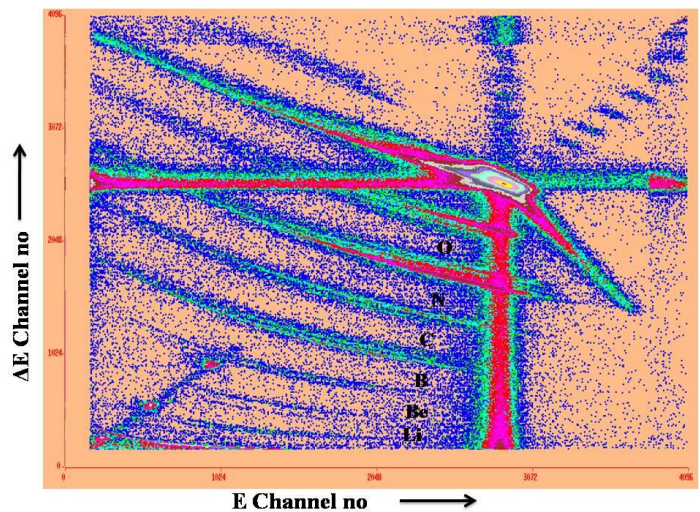

Figure 12: Typical two dimensional spectrum of fragments obtained in the reaction $^{20}$Ne+$^{112}$Sn.